# Statistical Approach for Predicting Factors of Mood Method for Object Oriented

Firas Jassim[1], Fawzi Altaani[2]

[1] Management Information Systems Department,
Irbid National University, Irbid, Jordan

[2] Management Information Systems Department,
Irbid National University, Irbid, Jordan

**Abstract**
Object oriented design is becoming more popular in software development and object oriented design metrics which is an essential part of software environment. The main goal in this paper is to predict factors of MOOD method for OO using a statistical approach. Therefore, linear regression model is used to find the relationship between factors of MOOD method and their influences on OO software measurements. Fortunately, through this process a prediction could be made for the line of code (*LOC)*, number of classes (NOC), number of methods (NOM), and number of attributes (NOA). These measurements permit designers to access the software early in process, making changes that will reduce complexity and improve the continuing capability of the design.

## 1. Introduction

Software metrics are most often proposed as the measurement tools of choice in empirical studies in software engineering, and the field of software metrics is the most often discussed from the perspective referred to as measurement theory. Software Metrics can be defined by measuring quality or characteristic of a software objects in any complex software project. Object oriented approach is capable of classifying the problem in terms of objects and provide many benefits like reliability, reusability, decomposition of problem into easily understood object and aiding of future modifications [2]. Nowadays, a quality engineer can choose from a large number of object–oriented metrics. The question posed is not the lack of metrics but the selection of those metrics which meet the specific needs of each software project. A quality engineer has to face the problem of selecting the appropriate set of metrics for his software measurements. A number of object–oriented metrics exploit the knowledge gained from metrics used in structured programming and adjust such measurements so as to satisfy the needs of object–oriented programming. On the other hand, other object–oriented metrics have been developed specifically for object–oriented programming and it would be pointless to apply them to structured programming [6]. Recently, many companies have started to introduce object-oriented (OO) technologies into their software development process. Many researchers have proposed several metrics suitable for measuring the size and the complexity of OO software. Some of them are in terms of Function Point (FP), others are in the terms of Lines of Code (LOC). Traditional metrics such as (FP) are unsatisfactory for predicting software size. On the other hand, LOC are quit satisfactory because it can be used to measure the software size [1, 7].

## 2. MOOD Method

The MOOD (Metrics for Object-Oriented Design) method is a collection of metrics which is used to evaluate the main abstraction of OO [4], such as *inheritance*, *encapsulation*, *coupling*, and *information hiding* or *polymorphism* and finally how to reuse that, together, for the increase in software quality. MOOD includes the following metrics [3, 5, 6, 13]:

- Method Hiding Factor (MHF)

- Attribute Hiding Factor (AHF)

- Method Inheritance Factor (MIF)

- Attribute Inheritance Factor (AIF)

- Coupling Factor (CF)

- Polymorphism Factor (PF)

These metrics are intended to presents the presence or the absence of a certain property or attribute. Mathematically speaking, it can be viewed as probabilities ranging from 0 (total absence) to 1 (total presence).

*Objects* are an encapsulation of information that is relative to some entity. The *class* can be viewed as an abstract data type (ADT), which includes two types of features: methods and attributes, where the number of defined methods in a class $C_i$ is given as:

$$M_d(C_i) = M_v(C_i) + M_h(C_i) \quad (1)$$

$M_d$ (represents defined methods), $M_v$ (represents visible methods), and $M_h$ (represents hidden methods).

Then we define the Method Hiding Factor (*MHF*), as follows:

$$MHF = \frac{\sum_{i=1}^{TC} M_h(C_i)}{\sum_{i=1}^{TC} M_d(C_i)}, \quad (2)$$

$Tc = Total\ Classes$

Conversely, the number of attributes defined in class $C_i$ (using the same manner above) is given by:

$$A_d(C_i) = A_v(C_i) + A_h(C_i) \quad (3)$$

$$AHF = \frac{\sum_{i=1}^{TC} A_h(C_i)}{\sum_{i=1}^{TC} A_d(C_i)} \quad (4)$$

And all other factors are calculating using similar mathematical formulas. So, *MIF* and *AIF* can be defined through equations (5) and (6), as:

$$MIF = \frac{\sum_{i=1}^{TC} M_i(C_i)}{\sum_{i=1}^{TC} M_d(C_i)} \quad (5)$$

AIF is defined as the ratio of the sum of inherited attributes in all classes of the system under consideration to the total number of available attributes (locally defined plus inherited) for all classes

$$AIF = \frac{\sum_{i=1}^{TC} A_i(C_i)}{\sum_{i=1}^{TC} A_d(C_i)} \quad (6)$$

PF is defined as the ratio of the actual number of possible different polymorphic situation for class $C_i$ to the maximum number of possible distinct polymorphic situations for class $C_i$, and can be defined as:

$$PF = \frac{\sum_{i=1}^{TC} M_o(C_i)}{\sum_{i=1}^{TC} [M_n \times DC(C_i)]} \quad (7)$$

where $M_o$ represents overridden methods, $M_n$ for new methods, and DC for descendants methods.

Polymorphism arises from inheritance and [10] suggest that in some cases overriding methods could contribute to reduce complexity and therefore to make the system more understandable and easier to maintain. While, [14] have shown that this metric is a valid measure within the context of the theoretical framework.

Finally, CF is defined as the ratio of the maximum possible number of couplings in the system to the actual number of couplings not imputable to inheritance.

$$CF = \frac{\sum_{i=1}^{TC} \left[ \sum_{j=1}^{TC} is\_client(C_i, C_j) \right]}{TC^2 - TC} \quad (8)$$

where:

$TC^2$-$TC$ = maximum number of coupling in a system with TC classes.

$$is\_client(C_i, C_j) = \begin{cases} 1 & iff\ C_i \Rightarrow C_j \wedge C_i \neq C_j \\ o & otherwise \end{cases} \quad (9)$$

Coupling Factor (CF) has a very high positive correlation with all quality measures [11]. Therefore, as coupling among classes increases, the defect density and normalized rework is also expected to increase. This result shows that coupling in software systems has a strong negative impact on software quality and then should be avoided during design. In fact, many authors have noted that it is desirable that classes communicate with as few others as possible because coupling relations increase complexity, reduce encapsulation and reuse.

## 3. Estimation of Factors

MOOD method used widely to measure many target OO programs and many studies have compare it with other methods. Mainly, our focus will be on line of code (LOC), number of classes (NOC), number of methods (NOM), and number of attributes (NOA), so to reach this; we have collect our data from 33 systems [9, 10, 12, 14] to be suitable for normal distribution curve[1]. Results obtained using SPSS package.

Table 1: Product metrics from 33 commercial samples

|   | NOL | NOC | NOM | NOA |
|---|---|---|---|---|
| 1 | 15837 | 65 | 1446 | 537 |
| 2 | 23570 | 57 | 1535 | 876 |
| 3 | 47106 | 91 | 2141 | 1178 |
| 4 | 23154 | 51 | 1420 | 538 |
| 5 | 20747 | 154 | 2814 | 1113 |
| 6 | 44930 | 92 | 2224 | 1132 |
| 7 | 28582 | 71 | 1978 | 839 |
| 8 | 19254 | 69 | 1815 | 675 |
| 9 | 20085 | 74 | 1876 | 700 |
| 10 | 57086 | 140 | 322 | 81 |
| 11 | 92231 | 201 | 481 | 124 |
| 12 | 167541 | 355 | 735 | 204 |
| 13 | 261260 | 562 | 1193 | 297 |
| 14 | 838128 | 1966 | 3227 | 611 |
| 15 | 2062982 | 5107 | 6735 | 2297 |
| 16 | 2129555 | 5035 | 7292 | 2294 |
| 17 | 1948354 | 4566 | 5975 | 2095 |
| 18 | 64492 | 222 | 210 | 81 |
| 19 | 70514 | 243 | 229 | 88 |
| 20 | 113919 | 349 | 325 | 132 |
| 21 | 177356 | 565 | 516 | 185 |
| 22 | 6593 | 324 | 1310 | 60 |
| 23 | 1023 | 25 | 103 | 220 |
| 24 | 1729 | 20 | 134 | 185 |
| 25 | 50000 | 46 | 2025 | 510 |
| 26 | 300000 | 1000 | 11000 | 10960 |
| 27 | 500000 | 1617 | 37191 | 17141 |
| 28 | 9189 | 339 | 1993 | 4022 |
| 29 | 7102 | 45 | 711 | 482 |
| 30 | 830 | 10 | 175 | 89 |
| 31 | 1602 | 26 | 180 | 247 |
| 32 | 3451 | 18 | 170 | 145 |
| 33 | 549 | 15 | 33 | 172 |
| Total N | 33 | 33 | 33 | 33 |

According to table (1), we can plot the relation between LOC (in the x-axis), and NOC, NOM, and NOA (in the y-axis), as shown in fig.1.

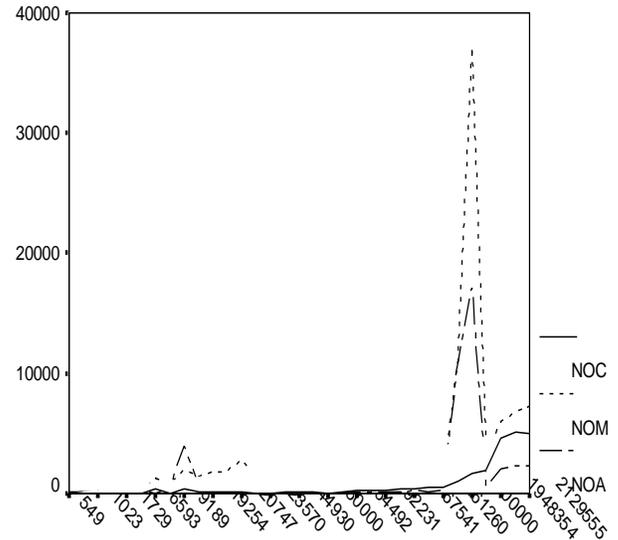

Fig. 1 The relationship between LOC and (NOC, NOM, and NOA)

Now, by implementing log transform to avoid large number scale we can plot the data again as fig. 2.

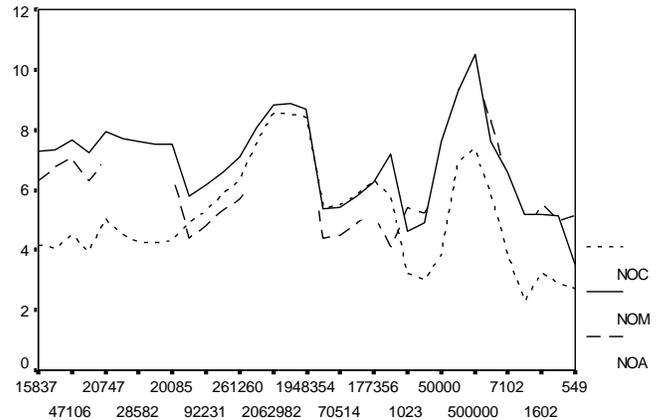

Fig. 2 The logarithmic relationship between LOC and (NOC, NOM, and NOA)

The main contribution in this article is to use statistics, especially regression; to predict number of classes needed for the software, also number of attributes and methods needed. Hence, linear regression model is used to find the relationship between factors and their influences on OO software measurements. Fortunately, through this process we can predict the suitable number of *LOC*, classes (objects), methods, and attributes we need to satisfy the software metrics using MOOD.

---

[1] Normal distribution needs more than thirty observation, while t distribution needs less than thirty observations, see [11].

## 4. Regression Analysis

Actually, we can use linear regression model to predict the *LOC*, *NOC*, *NOM*, and *NOA* needed. Statistically speaking, In order to investigate the correlations and relationships between the object-oriented metrics and software quality we conducted a correlation and a multiple linear regression analysis. The mathematical formula for the model is as follows:

$$LOC = \beta_0 + \beta_1\, NOC + \beta_2\, NOM + \beta_3 NOA \quad (10)$$

$$NOC = \beta_0 + \beta_1\, LOC + \beta_2\, NOM + \beta_3 NOA \quad (11)$$

$$NOM = \beta_0 + \beta_1\, LOC + \beta_2 NOC + \beta_3\, NOA \quad (12)$$

$$NOA = \beta_0 + \beta_1\, LOC + \beta_2 NOC + \beta_3\, NOM \quad (13)$$

Each time we have used one variable as an independent variable while the others as the dependent variables. To reach the fact that, each one of these variables responsible for the efficiency of the MOOD method. The regression analysis shows the values of the coefficients of the model ($\beta_0, \beta_1, \beta_2,$ and $\beta_3$).

The independent variable in an experiment is the variable that is systematically manipulated by the investigator. In most experiments, the investigator is interested in determining the effect that one variable; has one or more effect on the other variables. On the other hand, the dependent variable in an experiment is the variable that the investigator measures to determine the effect of the independent variable.

First, we consider LOC as the dependent variable and the other factors as the independent variables, equation 10, table (2) shows the value of ($\beta_0, \beta_1, \beta_2,$ and $\beta_3$), and the significances (p-value).

Table 2: Results of $\beta_0, \beta_1, \beta_2,$ and $\beta_3$ when LOC is the dependent variable

| | Regression coefficients | p-value |
|---|---|---|
| $\beta_0$ | -9458.918 | 0.220 |
| $\beta_1$ | 421.994 | 0.000 |
| $\beta_2$ | 3.025 | 0.327 |
| $\beta_3$ | -16.009 | 0.008 |

So, if we want to use the values of the coefficients above, we may re-write the regression line as:

LOC = -9458.918 + 421.994 NOC + 3.025 NOM - 16.009 NOA

Therefore, if we want to predict the value of *LOC* we can substitute the given values of *NOC*, *NOM*, and *NOA* in the above formula and get an estimated (predicted) value for *LOC*. Also, from the values of p-value we can see that the values of ($\beta_1$ and $\beta_3$) only are less than 0.05, so we can conclude that *LOC* are mainly affected by *NOC* and *NOA*. On the other hand, *NOM* does not affect *LOC* too much.

There is some statistical measures used to measure the goodness of fit and it is an indicator of how well the model fits the data. The higher the value of R square, the more accurate the model is. These values can be seen in table (3).

Table 3: The value of R square and adjusted R square for the regression model

| Model | R | R Square | Adjusted R Square | Std. Error of the Estimate |
|---|---|---|---|---|
| 1 | .998[a] | .996 | .996 | 37024.69 |

Since the value of significant (p-value) is less than 0.05. This means that *LOC* mainly affect the other factor according to table (4), which shows the **ANOVA** (**AN**alysis **O**f **VA**riance).

Table 4: ANOVA results for LOC as the dependent variable

| Model | | Sum of Squares | df | Mean Square | F | Sig. |
|---|---|---|---|---|---|---|
| 1 | Regression | 1.12E+13 | 3 | 3.749E+12 | 2734.947 | .000[a] |
| | Residual | 3.98E+10 | 29 | 1370827508 | | |
| | Total | 1.13E+13 | 32 | | | |

[a]. Predictors: (Constant), NOA, NOC, NOM
[b]. Dependent Variable: NOL

Second, we consider *NOC* as the dependent variable and the other factors as the independent variables, table (5) shows the value of ($\beta_0, \beta_1, \beta_2,$ and $\beta_3$), and the significances (p-value).

Table 5: Results of $\beta_0, \beta_1, \beta_2,$ and $\beta_3$ when NOC is the dependent variable

| | Regression coefficients | p-value |
|---|---|---|
| $\beta_0$ | 24.439 | 0.179 |
| $\beta_1$ | 0.002 | 0.000 |
| $\beta_2$ | -0.006 | 0.397 |
| $\beta_3$ | 0.037 | 0.011 |

Also, if we want to use the values of the coefficients above, we may re-write the regression line as:

NOC = 24.439 + 0.002 LOC - 0.006 NOM + 0.037 NOA

Therefore, if we want to predict the value of *NOC* we can substitute the given values of *LOC*, *NOM*, and *NOA* in the above formula and get an estimated (predicted) value for *NOC*. Also, from the values of p-value we can see that the values of ($\beta_1$ and $\beta_3$) only are less than 0.05, so we can conclude that *NOC* are mainly affected by *LOC* and *NOA*. On the other hand, *NOM* does not affect *LOC* too much. As previously mentioned the values of R square and the ANOVA table are shown in tables 6 &7.

Table 6: The value of R square and adjusted R square for the regression model when NOC is the dependent variable

| Model | R | R Square | Adjusted R Square | Std. Error of the Estimate |
|---|---|---|---|---|
| 1 | .998[a] | .997 | .996 | 87.55 |

Table 7: ANOVA results for NOC as the dependent variable

| Model | | Sum of Squares | df | Mean Square | F | Sig. |
|---|---|---|---|---|---|---|
| 1 | Regression | 64118680 | 3 | 21372893.45 | 2788.435 | .000[a] |
| | Residual | 222280.2 | 29 | 7664.835 | | |
| | Total | 64340961 | 32 | | | |

a. Predictors: (Constant), NOA, NOL, NOM
b. Dependent Variable: NOC

Similarly, we can do the same thing for *NOM* and *NOA*, put we mainly focused on the *LOC* and *NOC* because of their main role in MOOD method [8].

## 5. Conclusions

A simple and easy technique has been constructed to use statistics for predicting the values of MOOD factors, in the same manner one can use this technique to estimate other factors rather than *LOC*, *NOC*, *NOM*, and *NOA*, which can be used to evaluate software quality. Additionally, using linear regression model can be extended to non-linear model and multivariate analysis to add more complicated model to give more accurate estimation for these factors and also use another statistical estimation approaches such as maximum Likelihood Estimator (MLE) to give better estimation than regression model, and to be standards for MOOD method and to give more accurate measurements for object-oriented metrics.

**Acknowledgments**

Author Firas A. Jassim pays his regards to Mrs. Hind Emad Qassim for giving her moral support and help to carry out this research work.

**Firas Jassim** received the BS degree in mathematics and computer applications from Al-Nahrain University, Baghdad, Iraq in 1997, and the MS degree in mathematics and computer applications from Al-Nahrain University, Baghdad, Iraq in 1999 and the PhD degree in computer information systems from the University of Banking and Financial Sciences, Amman, Jordan in 2012. His research interests are Image processing, image compression, image enhancement, image interpolation and simulation.

**Fawzi Altaani** received the BS degree in public administration from Al-Yermouk University, Irbid, Jordan 1990, and Higher Diploma in health service administration from university of Jordan, 1991 and the MS degree in health administration from Red Sea University, Soudan in 2004 and the PhD degree in managment information systems from the University of Banking and Financial Sciences, Amman, Jordan in 2010. His research interests are management information system and public administration, and Image processing.